\documentclass[twocolumn, prl,showpacs, preprintnumbers,superscriptaddress]{revtex4}
\usepackage{graphicx}
\newcommand{\be}{\begin{equation}}
\newcommand{\ee}{\end{equation}}
\newcommand{\bphi}{\mbox{\boldmath $\phi$}}
\newcommand{\bgamma}{\mbox{\boldmath $\gamma$}}

\newcommand{\rh}{R}
\font\mybb=msbm10 at 11pt

\def\bb#1{\hbox{\mybb#1}}

\def\ben{\begin{equation}}
\def\een{\end{equation}}
\def\bea{\begin{eqnarray}}
\def\eea{\end{eqnarray}}
\begin{document}
\title{Hopf solitons and elastic rods}
\author{Derek Harland}
\email{d.g.harland@durham.ac.uk}
\affiliation{Department of Mathematical Sciences, Durham University, Durham DH1 3LE, U.K.}
\author{Martin Speight}
\email{j.m.speight@leeds.ac.uk}
\affiliation{School of Mathematics, University of Leeds, Leeds LS2 9JT, U.K.}
\author{Paul Sutcliffe}
\email{p.m.sutcliffe@durham.ac.uk}
\affiliation{Department of Mathematical Sciences, Durham University, Durham DH1 3LE, U.K.}
\preprint{DCPT-10/49}
\date{October 2010}
\begin{abstract}
Hopf solitons in the Skyrme-Faddeev model are string-like topological
solitons classified by the integer-valued Hopf charge. 
In this paper we introduce an approximate description of Hopf solitons
in terms of elastic rods. The general form of the elastic rod energy
is derived from the field theory energy and is found to be an extension
of the classical Kirchhoff rod energy. Using a minimal extension of
the Kirchhoff energy, it is shown that a simple elastic rod model 
can reproduce many of
the qualitative features of Hopf solitons in the Skyrme-Faddeev model.
Features that are captured by the model include the buckling
of the charge three solution, the formation of links at charges five and six,
and the minimal energy trefoil knot at charge seven.
\end{abstract}
\maketitle

%\section{Introduction}\news
The Skyrme-Faddeev model \cite{Fa2} is a field theory in three-dimensional
space. It possesses finite energy topological soliton solutions that
have a novel string-like structure and are known as Hopf solitons, as the 
topological classification is via the integer-valued Hopf charge $Q.$ 
This is a linking number, rather than the more familiar degree 
that appears in most theories with topological solitons.
 
It has been suggested \cite{FN2} that the model may be of relevance for a low
energy description of QCD, where the solitonic strings could describe
glueballs. Other applications have also been
proposed in the context of condensed matter physics, 
regarding the study of two charged condensates \cite{BFN}.

Substantial numerical work \cite{FN,GH,BS5,HS,Su} 
has produced a comprehensive catalogue of minimal energy solitons in the
Skyrme-Faddeev model. 
For $Q=1$ and $Q=2$ the minimal energy solitons are axially 
symmetric, but for larger values of $Q$ they
are more exotic and include knots and links.
These solutions have been obtained using sophisticated,
and computationally intensive, numerical simulations of the 
highly nonlinear field theory. The aim in this paper is to provide
a simple approximate description of Hopf solitons in terms of the
string core, thereby reducing the three-dimensional field theory
to an effective one-dimensional description. The resulting theory is
that of a generalized elastic rod, extending that of the classical
Kirchhoff rod. It is shown that a minimal extension provides a simple
model that is already capable of reproducing many of the qualitative
features of Hopf solitons.  

%\section{Hopf solitons in the Skyrme-Faddeev model}\news
The Skyrme-Faddeev field is a map $\bphi:
\bb{R}^3\mapsto S^2,$ which is realized as a three-component
unit vector $\bphi=(\phi_1,\phi_2,\phi_3).$ 
As this paper is concerned only with static
solutions then the model can be defined by its energy
 \be
E^{\rm SF}=
\int \partial_i\bphi\cdot\partial_i\bphi
+\frac{1}{2}(\partial_i\bphi\times\partial_j\bphi)^2
\ d^3x. \label{sfenergy}
\ee 
The first term in the energy is that of the usual $O(3)$ sigma model and
the second is a Skyrme term, required to provide a balance under
scaling and hence allow solitons with a finite non-zero size.

Finite energy boundary conditions require that the field tends to a
constant value at spatial infinity, which is chosen to be
$\bphi(\infty)=(0,0,1).$  This boundary condition
compactifies space to $S^3,$ so that the field becomes a map $\bphi:
S^3\mapsto S^2.$ Such maps are classified by $\pi_3(S^2)=\bb{Z},$ so
there is an integer-valued topological charge $Q,$ the Hopf charge.
It has a geometrical interpretation
as the linking number of two closed curves obtained as the preimages
of any two distinct points on the target two-sphere. 
A natural definition of the position of the soliton is provided 
by the preimage curve of the point $\bphi=(0,0,-1),$ which is 
antipodal to the vacuum value. The position of a Hopf soliton
is therefore a closed string, or possibly a collection of closed
strings since the preimage of any point may contain disconnected
components. This is the novel string-like aspect of Hopf solitons.
An energy bound of the form 
$E\ge a\, Q^{3/4},$ has been proved \cite{VK}, where
$a$ is a constant, and it is known that the fractional power is optimal
\cite{LY}. 
The sublinear growth has a simple physical explanation,
in that Hopf charge can be accrued by the linking of distinct components
of the position string (or by self-linking of a single component), so
that the total Hopf charge can be greater than the naive sum of its
components.  
 
Generically, the cross-section through
the string position locally resembles a 
planar soliton of the $O(3)$ sigma model.
The planar soliton has an internal phase,
 which may vary along the length of the curve to provide a twist. 
The simplest axially symmetric Hopf soliton of charge $Q$
may be pictured as a circular string, with a constant twist rate,
in which the
total change in the phase angle around the circle is $2\pi Q.$    
Such solutions exist for all $Q,$ but their energy
grows linearly with large $Q,$ in contrast to the $Q^{3/4}$ growth
of the minimal energy solitons. 
Only for $Q=1$ and $Q=2$ are the minimal energy solitons of this 
axial form. For $Q>2$ there is a buckling instability and
in fact the minimal energy soliton with $Q=3$ 
has a buckled conformation.

The $Q=4$ minimal energy soliton is axially symmetric
but is anomalous in that the cross-section is described
by a double planar soliton, with total phase twist $4\pi.$
 There is a local minimum of the energy that describes a buckled
$Q=4$ solution but it has an energy
that is a few percent above that of the global minimum.
There is a further solution at $Q=4$ that consists of two
 $Q=1$ solitons that are linked once.
If one links once a ring
of charge $Q_1$ and a ring of charge $Q_2$, the resulting configuration has
total charge $Q=Q_1+Q_2+2$ due to
the extra linking of the preimage curves of the components.
The charge four link has an energy between that of the buckled solution
and the global minimum. 

Minimal energy solitons with $Q=5$ and $Q=6$ are both links, with
two components and a single linking.
In the first case the components have 
charges one and two,
and in the second case both components have 
charge two.
The first minimal energy knot appears at $Q=7.$
It takes the form of a trefoil knot, which has crossing number three,
and has four units of twist, to make the total Hopf charge equal to seven.
 
\begin{figure}[ht]
\begin{center}
\includegraphics[width=8cm]{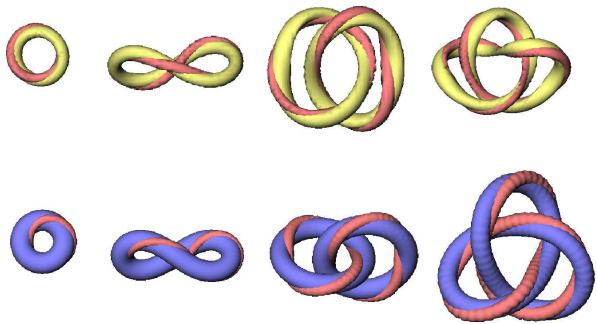}
\caption{{\em Top row:} The string position (yellow tubes) for minimal energy
Hopf solitons in the Skyrme-Faddeev model for charges $Q=1,3,6,7.$
The red curves indicate the twist. \break
{\em Bottom row:} The corresponding minimal energy elastic rods 
(blue tubes). To aid visualization,
the displayed rod thickness is only half the true thickness.
 }
\label{fig-compare}
\end{center}
\end{figure}

 For later comparison, a selection of
minimal energy Hopf solitons are displayed in the top row of 
Figure~\ref{fig-compare}. For clarity the string position is 
displayed by plotting a tube around the string, given by an isosurface
where $\phi_3=-0.8.$ The red curve indicates the twist and is
plotted in a similar fashion from the preimage of a point close
to the vacuum value.
%\section{An elastic rod energy}\news

The general form of an elastic rod energy
will now be derived 
from the Skyrme-Faddeev field theory energy (\ref{sfenergy}).
The approach taken is to define a field configuration by specifying
a closed curve $\bgamma(s)\in \bb{R}^3,$ where $s\in[0,L]$ 
is an arclength parameter, together with a quasiperiodic twist function 
$\alpha(s)$ such that $\alpha(L)=\alpha(0)+2\pi m,$ where $m\in \bb{Z}.$
The field configuration is defined by introducing tubular coordinates
centered on the curve and then mapping the tube to the target two-sphere
by first twisting each transverse disk using $\alpha(s),$ and then applying
a fixed degree one map from the disk to $S^2.$ Physically, the curve
may be pictured as the location of the string, the fixed degree one map
is the analogue of the planar soliton and the twist function specifies 
its internal phase.

To make the above construction explicit, let $\kappa(s)$ 
and $\tau(s)$  
denote the curvature and torsion of the curve $\bgamma(s)$ and consider the
standard Frenet frame with normal ${\bf N}(s)$ and 
binormal ${\bf B}(s).$ Take the radius of the transverse disk to
be equal to the radius of curvature $1/\kappa(s).$ 
The relation between the Euclidean coordinates ${\bf x}\in\bb{R}^3$
and the tubular coordinates $s,\rho,\theta$ is
\be
{\bf x}=\bgamma(s)+\frac{\rho}{\kappa}
({\bf N}(s)\cos\theta + {\bf B}(s)\sin\theta),
\label{tubular}
\ee
where $\rho$ and $\theta$ are polar coordinates in
the unit disk. 

The degree one map from the disk to $S^2$ 
has the form 
\be
\bphi=(\sin f \cos\theta, \sin f \sin\theta, \cos f),
\label{hh}
\ee
where $f(\rho)$ is a profile function satisfying the
boundary conditions $f(0)=\pi$ and $f(1)=0.$ 

Assuming the tubular region has no self-intersections, then for 
points in $\bb{R}^3$ inside the tubular region  
the above construction provides an explicit field $\bphi,$
and outside this region the field is set to its vacuum value
$\bphi=(0,0,1).$ This defines a field $\bphi$ throughout
$\bb{R}^3$ with a well-defined Hopf charge. In the case that
the curve $\bgamma$ is an unknot then the Hopf charge is
equal to the number of revolutions of the twist function, $Q=m.$

Substituting this ansatz into the Skyrme-Faddeev energy
(\ref{sfenergy}) gives, after a lengthy calculation, the
energy
\bea
E=\int_0^L \bigg\{
b_1+c_1\kappa^2&+&(\alpha'-\tau)^2\bigg(\frac{b_2}{\kappa^2}+c_2\bigg)
\nonumber\\
&+&
\frac{\kappa'^2}{\kappa^4}\bigg(b_3+c_3\kappa^2\bigg)
\bigg\}\, ds,
\label{roden}
\eea
where the constants are given in terms of integrals of
the profile function and its derivative.
The terms in the energy associated with the coefficients $b_1,b_2,b_3$ derive
from the sigma model term and the
terms associated with the coefficients $c_1,c_2,c_3$ are obtained
from the Skyrme-term.

The energy (\ref{roden}) is a function of the position curve
$\bgamma(s)$ and the twist function $\alpha(s),$ and 
is a natural generalization of the classical Kirchhoff energy for elastic
rods. The Kirchhoff energy is recovered if all coefficients except
$c_1$ and $c_2$ are set to zero. In this case, the two terms 
correspond to a bending and a twisting energy and  
the rod length $L$ needs to be fixed.  

The main aim
of the present paper is to show that an elastic rod model can reproduce
many of the qualitative features of Hopf solitons. This motivates 
a limited phenomenological approach, in which a simplified form
of the energy (\ref{roden}) is considered, by setting some of the coefficients
to zero. The remaining coefficients are then fixed by fitting to selected
properties of Hopf solitons. 

The starting point is the Kirchhoff energy, therefore $c_1$ and $c_2$
are taken to be non-zero. To balance the scaling properties of these
two terms requires at least
one contribution that is obtained from the sigma model term. We choose
 the simplest possibility, namely to take $b_1$ to be non-zero, so that
there is a contribution to the rod energy that simply
involves its length $L.$ The reduced elastic rod energy is therefore
defined by setting $b_2=b_3=c_3=0.$ In particular, this means that 
terms involving the derivative of the curvature are ignored in the
reduced theory.
 
%\section{Minimal energy rods}\news
The reduced elastic rod energy contains only three parameters. Two of these
parameters simply define the energy and length units of the model, which can
be fixed by matching to the energy and length of the $Q=1$ Hopf soliton.
This means that, in suitable units, two of the three coefficients may be
set to unity and the energy of the reduced elastic rod model can be
taken to be
\be
E=\int_0^L\bigg(
1 +\kappa^2+C(\alpha'-\tau)^2
\bigg)\,ds.
\label{ren}
\ee
A rod that is a minimizer of the energy (\ref{ren}) is in fact a Kirchhoff rod,
as it is also a minimizer of the Kirchhoff energy.
The additional feature induced by the first term in (\ref{ren}) is
that the rod length $L$ is not fixed, but is itself determined by
energy minimization. 

Axially symmetric Hopf solitons of charge $Q$ are modelled by
circular rods with a linear twist function $\alpha=2\pi Qs/L.$
They have energies and lengths given by 
\be 
E_Q^{\rm O}=4\pi\sqrt{1+CQ^2}, \quad\quad
L_Q^{\rm O}=2\pi\sqrt{1+CQ^2}.
\ee
Note that this energy formula captures the linear growth for
large $Q$
found for axially symmetric Hopf solitons.

As shown by Michell \cite{Mi} in 1889 , a circular rod will buckle if the total
twist $2\pi Q$ exceeds a critical value 
$
2\pi Q>2\pi\sqrt{3}/C.
$
For the first buckling instability of an axial rod to occur at $Q=3$ 
requires that $\frac{1}{\sqrt{3}}<C<\frac{\sqrt{3}}{2}.$
The analysis below reveals that the rod energies underestimate the 
soliton energies, so twisting is not sufficiently penalized.
To minimize this deficiency $C$ should be taken to be
close to its upper limit, therefore in the following we set $C=0.85.$
With this value of $C$ the above formulae yield
$
E_1^{\rm O}=17.09, \ L_1^{\rm O}=8.55, \
E_2^{\rm O}=26.36, \ L_2^{\rm O}=13.18, \
E_3^{\rm O}=36.96, \ L_3^{\rm O}=18.48.
$
A comparison of the $Q=1$ and $Q=2$ rods reveals that
$E_2^{\rm O}/E_1^{\rm O}=L_2^{\rm O}/L_1^{\rm O}=1.54.$
For Hopf solitons these ratios are
$E_2^{\rm SF}/E_1^{\rm SF}=1.63$ and $L_2^{\rm SF}/L_1^{\rm SF}=1.45,$
so the rod model is reasonably accurate, with an error of around
$6\%$ for both the energy and length of the $Q=2$ solution. 
The $Q=1$ axial rod is displayed as the first image in the bottom row
of Figure~\ref{fig-compare}.

To avoid self-intersection of the rod we impose an additional
constraint by assigning a radius, $\rh,$ to the cross-section 
of the rod and demand that this thickened 
rod does not self-intersect.
For the rod thickness to have no influence on the axial
solutions requires that $\rh\le L_1/(2\pi)=1.36.$ 
To maximize rod energies, we set $\rh$ equal to this upper limit.

To numerically compute minimal energy rods the curve $\bgamma$
is discretized into a polygonal curve with 100 vertices.
A simulated annealing method is used to minimise a discrete
version of the energy (\ref{ren}), with a condition of 
equal edge lengths imposed using a penalty function.
A discrete version of curvature is calculated 
from the angle between two neighbouring edges \cite{rawdon} and
the self-intersection of the rod is excluded by imposing 
an upper limit on the curvature $\kappa\le 1/R,$ and a lower limit on
the separation between specific subsets of pairs of vertices,
using the algorithm presented in \cite{rawdon}.
The twist contribution to the energy is 
evaluated by identifying the combination $\alpha'-\tau$ with the
rate of change of the angle between the material frame of the rod
and the twist free Bishop frame, as described in \cite{LS}.

Our numerical computations determine the energies of the 
first two buckled rods to be
$
E_3^\infty=35.35,$ and $
E_4^\infty=44.16.
$
The buckled minimal energy $Q=3$ rod is displayed as the second image
in the bottom row of Figure~\ref{fig-compare}, where it can be seen
that it closely resembles the corresponding Hopf soliton.

Links can be formed by putting together two rods, provided their radii 
are first increased to allow a sufficient gap to accommodate the 
other component of the link. 
Let $E_Q^\odot$  denote the energy of the axial rod with a radius
at least $2\rh,$ so that there is enough room to slip another rod 
through its centre.
Note $L_3^{\rm O}>2L_1^{\rm O},$ hence
for $Q\ge 3$ the axial rod is already large enough to insert another
rod giving $E_Q^\odot=E_Q^{\rm O}.$ However, for $Q=1$ and $Q=2$ the
natural radius must be increased to give
\be
E_Q^\odot
=\frac{\pi}{\rh}(4\rh^2+1+CQ^2).
\ee
With the values of $C$ and $\rh$ given earlier this formula results
in the energies  
$
E_1^\odot=21.36,\
E_2^\odot=27.25.
$
Forming a single link between a rod of charge $Q_1$ and 
a rod of charge $Q_2$ produces a link of charge $Q=Q_1+Q_2+2.$
We denote the energy of such a linked rod configuration by
$E_Q^{Q_1,Q_2}.$ If both $Q_1$ and $Q_2$ are less than three
then both components of the link will be axial rods and the
energy is simply obtained by addition as
$E_Q^{Q_1,Q_2}=E_{Q_1}^\odot+E_{Q_2}^\odot.$
This gives the link energies
$ E_4^{1,1}=2E_1^\odot=42.73, \
 E_5^{2,1}=E_2^\odot+E_1^\odot=48.62, \
E_6^{2,2}=2E_2^\odot=54.51.$
Thus for $Q=4$ the link is slightly lower in energy than the 
buckled rod, which agrees with the result for Hopf solitons, where
solutions of both types also exist.
Recall that the minimal energy Hopf soliton for $Q=4$ 
is anomalous, in that its cross-section is a double planar soliton, 
therefore in its current form the elastic rod model is unable to describe
this soliton.
It is possible to generalize the rod model to a cross-section with
arbitrary degree, 
 but this refinement is more appropriate for
future investigations that include all the possible terms in the 
energy.
\begin{figure}[ht]
\begin{center}
\includegraphics[width=8cm]{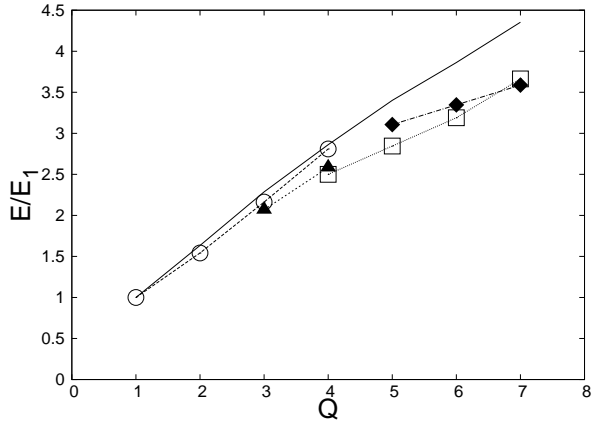}
\caption{
The data points are (normalized) elastic rod energies: 
circles for circular rods, triangles for buckled rods, squares for links,
and diamonds for knots.
The upper solid curve is the (normalized) minimal energy 
of Hopf solitons in the Skyrme-Faddeev model.
}
\label{fig-en}
\end{center}
\end{figure}
A lower bound on the energy of a 
link formed from rods of charge three and two is
$ E_7^{3,2}\ge E_3^\infty+E_2^\odot=62.60. $
The energies of trefoil knots with charge $Q$ are denoted 
by $E_Q^K,$ and have $Q-3$ turns, with a contribution of three to the Hopf
charge due to the self-linking of the trefoil knot. 
Our numerical computations provide the following 
trefoil knot energies 
$
E_5^K=53.09,\
E_6^K=57.20,\
E_7^K=61.32.
$

Comparing the above link and knot energies shows that links are preferred
for $Q=5$ and $Q=6,$ but at $Q=7$ the knot energy is below the lower bound
of the link. These results show that the forms of minimal energy rods 
for $Q=5,6,7$ are in agreement with those of minimal energy Hopf solitons.
The final two plots displayed in the bottom row of Figure~\ref{fig-compare}
are the minimal energy $Q=6$ and $Q=7$ rods, whose forms compare reasonably
well with the corresponding 
minimal energy Hopf solitons presented in the top row. 

The data points in Figure~\ref{fig-en} are the various rod energies, 
normalized by the energy $E_1^{\rm O}$ of the $Q=1$ rod.
For comparison, the upper solid curve is the minimal energy 
in the Skyrme-Faddeev model, normalized by the energy $E_1^{\rm SF}$ 
of the $Q=1$ Hopf soliton.
Although the rod energies grow too slowly in comparison to the 
soliton energies, we have seen that
several qualitative features are correctly reproduced regarding the 
minimal energy configurations, that is,
axial for $Q=1,2$, buckled for $Q=3,$ links for $Q=5,6$ and the first 
minimal knot at $Q=7.$ 

It is perhaps not surprising that the rod energies underestimate the 
Hopf soliton energies, because the non-local interaction between different
parts of the rod is very simplistic and does not provide a direct contribution
to the energy.
It might be possible to produce a more accurate model by introducing
a more sophisticated non-local interaction, but then some of the simplicity
and elegance of the current model would be lost. Similarly, the extra terms
neglected to produce the reduced rod model (for example, terms involving
the derivative of the curvature) could also be considered and will again 
lead to increases in rod energies. 
This may provide a more accurate quantitative rod model, 
and could be investigated in the future.

%\section{Conclusion}\news
It is certainly of interest to obtain an improved
quantitative description of Hopf solitons, to build on the successful
qualitative results presented here. An accurate rod approximation would
be a very useful tool for investigating Hopf soliton 
issues that are currently difficult to
study within the field theory, such as the existence of non-torus knots
and the structure of solitons for large Hopf charges. 

Finally, note that if
the rod energy simply consisted of the first term in (\ref{ren})
then our problem would coincide with the construction of 
ideal knots and links \cite{ideal}, in which the energy function
is the length of the rod for a fixed thickness. The energy (\ref{ren}) is 
therefore an interesting hybrid of the Kirchhoff elastic rod
and ideal knot energies.

\begin{acknowledgments}
Many thanks to Alex Collins and Dave Foster for useful discussions.
We acknowledge the EPSRC funding EP/G038775/1
and EP/G009678/1.
\end{acknowledgments}

\end{document}